\documentclass[10pt]{article}
\usepackage{amssymb}
\usepackage{amsmath}

        \newcommand{\mbra}[1]{\langle {#1}|}
                 \newcommand{\mket}[1]{ \left|{#1} \right\rangle }
        \newcommand{\sks}[2]{\langle {#1}|{#2}\rangle }

        \newcommand{\beq}{\begin{eqnarray}}
        \newcommand{\eeq}{\end{eqnarray}}

\newcommand{\lbl}[1]{\label{#1}}

 \linethickness{0.4pt}
\begin{document}
\newtheorem {proposition}{Proposition}[section]
\newtheorem{lemma}{Lemma}[section]
\newtheorem{theorem}{Theorem}[section]
\newtheorem{corollary}{Corollary}[section]
\begin{flushright}
                IFT UwB /14/2002 \\
                

\end{flushright}
\today
\bigskip
\bigskip
\begin{center}
{\Large\bf  Systems with  Intensity--Dependent Conversion
  Integrable by Finite Orthogonal Polynomials}

\end{center}
\bigskip
\bigskip
\begin{center}
{\bf  Maciej Horowski*, Goce Chadzitaskos**,\\  Anatol
Odzijewicz*, Agnieszka Tereszkiewicz*} \footnote{ E-mail:
*aodzijew@labfiz.uwb.edu.pl, horowski@alpha.uwb.edu.pl,
tereszk@alpha.uwb.edu.pl, **goce.chadzitaskos@fjfi.cvut.cz}
\end{center}
\bigskip
\bigskip
\begin{center}
{*Institute of Theoretical Physics University in Bia{\l}ystok, \\Lipowa 41, 15-424
Bia{\l}ystok,
Poland\\
\medskip
{**}Depart. of Physics, FNSPE, Czech Technical University,\\
Brehova 7, CZ - 115 19 Praha 1}
\end{center}
\bigskip\bigskip

\begin{abstract}
\noindent
  We present exact solutions of a class of the nonlinear models which describe
  the parametric conversion of photons.
   Hamiltonians of these models are related to the
  classes of finite orthogonal polynomials. The spectra and
exact expressions for eigenvectors of these Hamiltonians are obtained.
\end{abstract}


\vspace{1.5cm}

\section{Introduction}
\renewcommand{\theequation}{1.\arabic{equation}}

In nonlinear optical models the influence of a medium on electromagnetic field
$\Big(\overrightarrow{E},\overrightarrow{B}\Big)$ is described by the material source-free Maxwell equations, where in
general a functional dependence $\overrightarrow{P}=\overrightarrow{P}\big[\overrightarrow{E}\big]$ of the polarization
$\overrightarrow{P}$ on the electric field $\overrightarrow{E}$ is assumed (see e.g.  \cite{P-L}, \cite{H-O-T 2},
\cite{B-C}). This dependence describes  complicated microstructure of the medium and the nonlinearity of the matter --
field interactions.
 Assuming the classical description of the medium,
 the field can be  quantized, and  the energy operator is obtained. In the case
of the two-mode field the operator is given by
  \beq\lbl{0.1}
\mathbf H=\mathbf H_0+e^{-i\mathbf H_0\;t}\;\mathbf H_I\;
e^{i\mathbf H_0\;t},
 \eeq
  where
  \beq\label{0.2}
\mathbf H_0=\omega_0\mathbf a_0^*\mathbf a_0+\omega_1\mathbf
a_1^*\mathbf a_1
 \eeq
  describes the free field, and the term
 \beq\label{0.3}
   \mathbf H_I=\sum\limits_{k,l,m,n=0}^\infty \Big(\alpha_{klmn}\left(\mathbf
a_0^*\right)^k\left(\mathbf a_1^*\right)^l \mathbf a_0^{m}\mathbf
a_1^{n}+h.c.\Big),
    \eeq
    where $\alpha_{1010}=\alpha_{0101}=0$,
     is the interaction Hamiltonian responsible for the light--matter interactions.
     The  annihilation and creation operators
     of two  modes $\mathbf  a_0,\mathbf a_1, \mathbf a_0^*, \mathbf
     a_1^*$    fulfill  the Heisenberg canonical commutation relations. Using the
     boson--number ordering, see \cite{O-H-T}, $\mathbf
     H_I$ can be expressed in the form:
  \beq\lbl{0.4}
   \mathbf H_I=\left(\sum\limits_{k_0,k_1=1}^\infty f_{k_0k_1}(\mathbf
a_0^*\mathbf a_0,\mathbf a_1^*\mathbf a_1)
   \mathbf a_0^{k_0}\mathbf a_1^{k_1}+h.c. \right)+\\
   +\left(\sum\limits_{k_0,k_1=0}^\infty g_{k_0k_1}(\mathbf a_0^*\mathbf a_0,\mathbf a_1^*\mathbf a_1)
   \mathbf a_0^{k_0}\left(\mathbf
   a_1^{*}\right)^{k_1}+h.c.\nonumber
   \right),
  \eeq
   where $f_{k_0k_1}(x,y)$ and $g_{k_0k_1}(x,y)$ are functions of  two
   arguments $x,y$. These functions are defined by the constants
   $\alpha_{klmn}$ and   are
   responsible for the light--matter interaction via the functional dependence
      $\overrightarrow{P}=\overrightarrow{P}[\overrightarrow{E}]$. If the sum in
   (\ref{0.3}) is finite then  $f_{k_0k_1}(x,y)$ and
   $g_{k_0k_1}(x,y)$ are polynomials. This case has been investigated
   during the last decade by many authors, see e.g. \cite{K},
   \cite{J}, \cite{K 1}. They   used  approximate or
   semiclassical methods.

       Let us give the interpretation of the particular
       terms of the Hamiltonian (\ref{0.4}). The term \newline
       $ g_{k_0k_1}(\mathbf a_0^*\mathbf a_0,\mathbf a_1^*\mathbf a_1)
   \mathbf a_0^{k_0}\left(\mathbf
   a_1^{*}\right)^{k_1}$ describes the process of simultaneous absorption of $k_0$ photons in
 mode $0$ and emission  of $k_1$ photons in mode $1.$ The
   probability of this process depends on $ g_{k_0k_1}(\mathbf a_0^*
   \mathbf a_0,\mathbf a_1^*\mathbf
   a_1)$, i.e. it depends on the intensity of light
   in the medium . So the function $ g_{k_0k_1}$ is a
   generalization of the coupling constant for the conversion.
Such a process is called the intensity dependent \cite{J}
   (or parametric \cite{Pe-Lu})  conversion of  $k_0$ photons in the
   mode $0$ into $k_1$ photons in mode $1$.
    The hermitian conjugate
   term $\big( g_{k_0k_1}(\mathbf a_0^*\mathbf a_0,\mathbf a_1^*\mathbf
   a_1)\mathbf a_0^{k_0}\left(\mathbf
   a_1^{*}\right)^{k_1}\big)^*$ describes the parametric conversion of $k_1$
   photons in mode $1$ into $k_0$ photons in  mode $0$
  with coupling given by the operator $\overline{g}_{k_0k_1}(\mathbf a_0^*\mathbf a_0+k_0,\mathbf a_1^*\mathbf
   a_1-k_1)$. By analogy the term $f_{k_0k_1}(\mathbf a_0^*\mathbf a_0,\mathbf a_1^*\mathbf
   a_1)\mathbf a_0^{k_0}\mathbf
   a_1^{k_1}$ corresponds to the process of absorption
   by the medium of the cluster consisting of the $k_0$ photons in
   the  mode $0$ and $k_1$ photons in the  mode $1$. The
   hermitian conjugate term describes the emission of the  same cluster
    by the medium.

In this paper we study the intensity dependent conversion of the fixed number $k_0$ of photons in  mode $0$
   into a fixed number $k_1$ of photons in mode $1$ and \textit{vice versa}.
   To   simplify the  notation let $h_0:=g_{00}+\overline{g}_{00}$  and
   $g:=g_{k_0k_1}$; the interaction Hamiltonian for such process takes the form
    \beq
\mathbf H_I=h_0\left( \mathbf a_0^{*}\mathbf a_0,\mathbf a_1^{*}
\mathbf a_1\right) + g\left( \mathbf a_0^{*}\mathbf a_0,\mathbf
a_1^{*}\mathbf a_1\right) \mathbf a_0 ^{k_0} \left(\mathbf a_1^{*}
\right)^{k_1}+\left( \mathbf a_0^{*}\right)^{k_0}\mathbf
a_1^{k_1}\overline{g}\left( \mathbf a_0^{*}\mathbf a_0, \mathbf
a_1^{*}\mathbf a_1\right)\lbl{Aa}
      \eeq
      with the complex--valued function
      \begin{equation}
 g\left(\mathbf a_0^{*}\mathbf a_0,\mathbf a_1^{*}\mathbf a_1\right) =
  e^{i \theta\left(\mathbf a_0^{*}\mathbf a_0,\mathbf a_1^{*}
 \mathbf a_1\right)}
 |g\left(\mathbf a_0^{*}\mathbf a_0,\mathbf a_1^{*}\mathbf a_1\right)|.
 \end{equation}

 This paper is a continuation of the program initiated in
   \cite{O-H-T}, \cite{H-O-T 2},  where the theory of
     orthogonal polynomials  was applied  to the solution of
   the eigenproblem for the Hamiltonians of the type (\ref{Aa}) for
   general functions $h_{0}$ and $g $. The   solution
   of the eigenproblem of the interaction Hamiltonian $\mathbf H_I$ is
   equivalent to the problem of the integrability of the system under
   consideration; it   follows from the fact that the solution of
   Schr\"{o}dinger equation for the Hamiltonian (\ref{0.1}) is
 \beq
      \mket{\psi(t)}=e^{-i\mathbf H_0t}e^{-i\mathbf
     H_It}\mket{\psi(0)}.\label{0.5}
       \eeq

 In Section 2 a family of operators commuting with the
Hamiltonians (\ref{0.2}) and  (\ref{0.3}) is found, i.e. the integrals of
 motion of the system. It is shown that the Hilbert space can be split into finite--dimensional
 subspaces invariant under the action of $\mathbf H_0$ and $\mathbf H_I$.
 Introducing the Fock basis in the subspaces, the matrix form of the interaction Hamiltonian
 is the Jacobi matrix.
 In Section 3 it is shown that the eigenproblem of the considered Hamiltonian  is equivalent
 to the moment problem of the theory of finite orthogonal
 polynomials. Using this equivalence we express the spectral decomposition of the
 Hamiltonian in terms of the orthogonal polynomials.
 Section 4 is devoted to the construction  of the families of Hamiltonians related to the fixed systems of orthonormal
 polynomials.
In Section 5 the families of integrable Hamiltonians are presented. We present examples of simplest Hamiltonians related
to the well known  families of finite orthogonal polynomials. The spectra and eigenvectors of these Hamiltonians are
presented.


\section{Conversion of fixed numbers of photons}
\renewcommand{\theequation}{2.\arabic{equation}}
\setcounter{equation}{0}

Let us show that the two mode Fock space $\mathcal H$   can be split
 into finite dimensional subspaces, invariant with   respect  to the
interaction Hamiltonian (\ref{Aa}) and the free Hamiltonian (\ref{0.2}),
 and therefore also invariant with  respect to the Hamiltonian (\ref{0.1}). First we observe that the
following operators
  \beq\lbl{1.1}
&\mathbf{K}:= k_1\mathbf a_0^*\mathbf a_0^{}+ k_0 \mathbf
a_1^*\mathbf a_1^{}&\\
&\mathbf R_\kappa:=\frac{k_\kappa-1}{2}+
   \sum\limits^{k_\kappa-1}_{s=1}\;
   \frac{\exp\left({-i{\frac{2\pi s }{k_\kappa}}\;\mathbf a_\kappa^{*}\mathbf
   a_\kappa}\right)}
   {\exp\left({i\frac{2\pi s}{k_\kappa}}\right)-1},\lbl{1.2}&
 \eeq
where $\kappa=0,1$, commute with $\mathbf H_0$ and $\mathbf H_I$, i.e. they are  integrals of motion of the system.
$k_0$ and $k_1$ are defined by (\ref{Aa}). The elements  of Fock basis of $\mathcal H$
       \begin{equation}
\left| n_0,n_1\right\rangle =\frac{1}{\sqrt{n_0!n_1!}%
}\left( \mathbf a_0^{*}\right) ^{n_0} \left( \mathbf
a_1^{*}\right) ^{n_1}\left| 0,0\right\rangle ,\;\;\;\;
n_0,n_1=0,1,\ldots\lbl{Aw}
\end{equation}
 are  eigenvectors of the operators (\ref{1.1}),
(\ref{1.2}):
           \beq
           &\mathbf K\;\mket{n_0,n_1} = (k_1n_0+k_0n_1 )\;\mket{n_0,n_1},&\\
&            \mathbf
R_\kappa\mket{n_0,n_1}=r_\kappa\mket{n_0,n_1},&
             \eeq
              where the eigenvalues $r_\kappa$ are equal to the remainder of
              the division of $n_\kappa$ by $k_\kappa$
               \big($r_\kappa=n_\kappa(\!\!\!\!\mod k_\kappa)$\big), see \cite{K}.
So one has the orthogonal decomposition
  \beq
                \mathcal{H}=\bigoplus_{\mu\in J}
 \mathcal{H}_{\mu} \label{0.13}
 \eeq
 of $\mathcal H$ into finite--dimensional Hilbert subspaces
 ${\mathcal H}_\mu$ labelled by multi--indices $ {\mu}:=(r_0,r_1,N)\in
J:=\{0,1,\ldots,k_0-1\}\times\{0,1,\ldots,k_1-1\}\times(\mathbb{N}\cup\{0\})$
and spanned by the vectors
 \beq
| n \rangle_{\mu}:=\big|\, r_0+k_0 n ,\;r_1+k_1(N-n)
\big\rangle\;\;\;n=0,1,\ldots,N
     \eeq
     of the Fock basis. The Hilbert subspace $\mathcal{H}_{\mu}$ is a common eigenspace
of the operators $\mathbf
 K,\mathbf R_0$ and $\mathbf R_1$ with dimension
 \beq
   \dim  \mathcal{H}_{\mu}=N+1.
 \eeq
Moreover, $N+1$ can be obtained as the only eigenvalue of the
dimension operator
   \beq
     \mathbf D=\frac{1}{k_0k_1}\mathbf K-\frac{1}{k_0}\mathbf
     R_0-\frac{1}{k_1}\mathbf R_1+1
    \eeq
    on $\mathcal{H}_\mu$.
According to the above let us introduce the operators

 \beq
 &\mathbf A_0:=\frac{1}{k_0}\mathbf a_0^{*}\mathbf a_0^{},&\\
 &\mathbf A:=g\left( \mathbf a_0^{*}\mathbf a_0,\mathbf
a_1^{*}\mathbf a_1\right) \mathbf a_0^{k_0} \left(\mathbf a_1^{*}
\right)^{k_1}.&
 \eeq
 and also replace the operators $\mathbf a^*_0\mathbf a_0$ and  $\mathbf a^*_1\mathbf a_1$ by $\mathbf
 A_0$ and $\mathbf K$.

  The operators $\mathbf A_0,\;\mathbf A,\;\mathbf A^*$ satisfy the
relations
 \beq
 \nonumber
& \left[ \mathbf A_0,\mathbf A\right] =-\mathbf A,\quad \left[
\mathbf A_0,\mathbf A^{*}\right] =\mathbf
A^{*}, &\\
&\mathbf A^{*}\mathbf A ={\cal G}\left( \mathbf
A_0-1,\mathbf{K}\right),  \lbl{Ass}&\\
& \mathbf A\mathbf A^{*} ={\cal G}\left( \mathbf
A_0,\mathbf{K}\right)  \nonumber,&
   \eeq
   where the function ${\cal G}\left( \mathbf
A_0,\mathbf{K}\right)  $ is
determined by $g$:
 \beq
 {\cal G}\left( \mathbf A_0,\mathbf{K}\right)&: =& \left|g
\left( k_0\mathbf A_0,\frac{1}{k_0}\mathbf{K}-k_1\mathbf A_0\right) \right| ^2(k_0 \mathbf A_0+1)(k_0 \mathbf
A_0+2)\ldots(k_0
\mathbf A_0+k_0)\times\nonumber\\
&\times&\left(\frac{1}{k_0}\mathbf{K}-k_1\mathbf
A_0\right)\left(\frac{1}{k_0}\mathbf{K}-k_1\mathbf
A_0-1\right)\ldots\left(\frac{1}{k_0}\mathbf{K}-k_1\mathbf
A_0-k_1+1\right) .\lbl{Ac}
 \eeq
  It can be shown that
\beq
 \mathbf A_0| n \rangle_{\mu}=\left(\frac{r_0}{k_0}+n\right)|
n \rangle_{\mu},
 \quad\quad\quad
 \mathbf A| n \rangle_{\mu}=b_{n-1,\mu}\mket{n-1}_\mu,
 \quad\quad\quad\mathbf A^*| n
 \rangle_{\mu}=\overline{b_{n,\mu}}\;\mket{n+1}_\mu, \lbl{Ab}
 \eeq
 where
  \beq\lbl{1.19}
b_{n,\mu}=e^{-i\phi\left(\frac{r_0}{k_0}+n,k_1r_0+k_0r_1+k_0k_1 N
\right)}\sqrt{{\cal G}\left(\frac{r_0}{k_0}+n,k_1r_0+k_0r_1+k_0k_1
N \right)},
 \eeq
 i.e. $\mathbf A_{0}$ is diagonal, $\mathbf A$ and $\mathbf
 A^{*}$ are weighted shift operators and
 $\phi$ denotes the phase factor of function $g$ expressed
in the new variables $n$ and $N.$ Let us notice that from  (\ref{Ac}) it follows that ${\cal G} \left( -1,N \right)=
{\cal G} \left( N,N \right)= 0$ what makes (\ref{Ab}) consistent.

The interaction Hamiltonian  (\ref{Aa}) now takes the form
\begin{equation}
\mathbf H_I= h\left( \mathbf A_0,\mathbf{K}\right) +\mathbf
A+\mathbf A^{*}  \lbl{Au} ,
\end{equation}
where $h$ is uniquely determined by $h_0$ and the free Hamiltonian $\mathbf H_0$ is given by
 \beq
 \mathbf H_0=(\omega_0k_0-\omega_1k_1)\mathbf A_0+\frac{\omega_1}{k_0}\mathbf
 K.
   \eeq
Thus the Hamiltonian describing our system belongs to the operator
algebra generated by the operators $\mathbf{K},\;\mathbf A_0,\;
\mathbf A,\; \mathbf A^*$\; ($\mathbf{K}$ commutes with the
others). The subspaces ${\mathcal H}_{\mu}$ are invariant
subspaces of the operators $\mathbf H_0, \;\mathbf H_I$ and
therefore of $\mathbf H$. Moreover, the action of these operators
on elements of the Fock basis is
 \beq
\mathbf H_0\,| n
\rangle_{\mu}=\Big(\,\omega_0r_0+\omega_1r_1+\omega_1k_1N+(\omega_0k_0-\omega_1k_1)n\,\Big)\,|n\rangle_{\mu}
   \eeq
   and
\begin{eqnarray}
\mathbf H_I\,| n \rangle_{\mu}= b_{n-1,\mu}\;| n-1
\rangle_{\mu}+a_{n,\mu}\;| n
\rangle_{\mu}\lbl{1.a}+\overline{b_{n,\mu}}\;| n+1
\rangle_{\mu},\lbl{23}
\end{eqnarray}
where $b_{n,\mu}$ is given by (\ref{1.19}), and
 \beq
\displaystyle{a_{n,\mu}=h\left(\frac{r_0}{k_0}+n,\;\;k_1r_0+k_0r_1+k_0k_1
N \right)}.
  \eeq
It follows from (\ref{1.a})  that the matrix form of the operator
$\mathbf H_{\mu}:=\mathbf H_I\left|_{{\mathcal H}_{\mu}}\right.$
in the basis $\big\{\mket{n}_\mu\big\}_{n=0}^N$ of
$\mathcal{H}_\mu$ is the Jacobi matrix (three diagonal and
hermitian matrix). This fact allows us to use the theory of
orthogonal polynomials to solve the eigenproblem of $\mathbf
H_{\mu}$ \cite{O-H-T}, \cite{H-O-T 2}.

\section{Integration and finite polynomials}
\renewcommand{\theequation}{3.\arabic{equation}}
\setcounter{equation}{0}

The three--term  formula (\ref{23}) suggests that the eigenproblem of $\mathbf H_\mu$ is strictly connected with the
theory of finite orthogonal polynomials. One can apply this theory under the additional assumption that $\mathbf H_\mu$
has $N+1$ different eigenvalues $\big\{ E_{l,\mu}\big\}_{l=0}^N$ with the corresponding eigenvectors
$\big\{\mket{E_{l,\mu}}\big\}_{l=0}^N,$ i.e.
\begin{eqnarray}
\mathbf H_\mu\mket{E_{l,\mu}}=
E_{l,\mu}\;\;\mket{E_{l,\mu}}.\lbl{Ba}
\end{eqnarray}
Then the spectral decomposition of the interaction Hamiltonian is
\begin{eqnarray} \label{spec}
\mathbf H_{I} = \sum_{\mu \in J} \sum_{l=0}^{N}
E_{l,\mu}\;\;\frac{\mket{E_{l,\mu}}\mbra{E_{l,\mu}}}{\sks{E_{l,\mu}}{E_{l,\mu}}}.
\end{eqnarray}

If we decompose $\mket{E_{l,\mu}}$  in the Fock basis
$\{|n\rangle_\mu\}_{n=0}^N$ of ${\mathcal H}_\mu$
\begin{eqnarray} \mket{E_{l,\mu}} = \sum_{n=0}^N
P_{n}^{\mu}\!(E_{l,\mu})\;\;|n\rangle_\mu, \lbl{Bb}
\end{eqnarray}
then from (\ref{23}), (\ref{Ba}) and (\ref{Bb}) it follows that the coefficients $P_{n}^{\mu}\!(E_{l,{{\mu}}})$ satisfy
the three term
 identity
 \beq
 E_{l,{{\mu}}}\;P_{n}^{\mu}\!(E_{l,{{\mu}}})  \lbl{24}
=\overline{b_{n-1,\mu}}\; \;P_{n-1}^\mu\!(E_{l,{{\mu}}})
 \displaystyle{+a_{n,\mu}\;P_{n}^\mu\!(E_{l,{{\mu}}})}
+b_{n,\mu}\;P_{n+1}^\mu\!(E_{l,{{\mu}}}),\;\;\;\;\;\;\;\;\;\; n,l=0,1,\ldots,N
 \eeq
 which can be considered as the recurrence relations
  for   $P_n^\mu\!(E_{l,\mu})$  with the initial condition
   $P_0^\mu \!(E_{l,\mu})\!\equiv~1.$ Thus $P_n^\mu\!(E_{l,\mu})$ is
   a polynomial  of degree $n$
    in the variable $E_{l,\mu}$. Since  $\mathbf H_\mu$ is hermitian, the set $\{ |E_{l,\mu}\rangle\}_{l=0}^{N}$
    forms an orthogonal basis in ${\mathcal H}_\mu$.
 The orthogonality relations for eigenvectors $\mket{E_{l,\mu}} $
 imply
the orthonormality relation in the set of polynomials $\{
P_n^\mu\!(E_{l,\mu})\}_{n=0}^{N}$:
\begin{eqnarray}
 \sum_{l=0}^{N}\;\overline{ P_n^\mu \!(E_{l,{{\mu}}})}\;P_m^\mu \!(E_{l,{{\mu}}})
 \frac{1}{\langle E_{l,{{\mu}}} | E_{l,{{\mu}}}
 \rangle} = \delta_{mn}~;\label{Bd}
 \end{eqnarray}
this allows to invert the formula (\ref{Bb}):
\begin{eqnarray} |n\rangle_\mu = \sum_{l=0}^N \frac{1}{\langle E_{l,\mu} | E_{l,\mu}
 \rangle}\overline{ P_{n}^{\mu}\!(E_{l,\mu})}\;\;|E_{l,\mu} \rangle. \lbl{Bc}
\end{eqnarray}

 In such a way we obtain a finite system of orthonormal polynomials  \;\;\;$\big\{
P^\mu_n\! \big\}_{n=0}^{N}$,  with respect to the weight function
 $\frac{1}{\sks{E_{l,\mu}}{E_{l,\mu}}}$, dependent on the discrete
 variables
$\big\{E_{l,\mu}\big\}_{l=0}^N$.

Since the interaction Hamiltonian $\mathbf H_{I}$ preserves the
decomposition (\ref{0.13}), the unitary one--parameter group
of Schr\"odinger evolution in the interaction picture $
e^{-i\mathbf H_I \,t} $ (see (\ref{0.5}))
 preserves the decomposition of the identity
 \beq \textbf{1}=
\sum_{\mu\in J} \sum_{n=0}^N|n\rangle
_{{{\mu}}}\;\;_{{{\mu}}}\langle n\,|\;.
  \eeq
 Thus, using the spectral decomposition (\ref{spec}) and the orthogonal relations (\ref{Bd})
  the operator $e^{-i\mathbf H_I  \,t}$ can be expressed in the form

\begin{eqnarray}
e^{-i \mathbf H_{I}t} = \sum_{\mu \in J} \sum_{l=0}^{N} e^{-i E_{l,\mu}
t}\;\;\frac{\mket{E_{l,\mu}}\mbra{E_{l,\mu}}}{\sks{E_{l,\mu}}{E_{l,\mu}}}.\lbl{evo}
\end{eqnarray}
Its matrix elements
        \beq \label{2.7}
       _{ \nu}\mbra{m} e^{-i\mathbf H_I \,t}\mket{n}_{{\mu}}=\delta_{NS}\;\delta_{r_0q_0}\delta_{r_1q_1 }\sum_{l=0}^N
\overline{P_{m}^\mu \!(E_{l,{{{\mu}}}})}\;P_{n}^\mu
\!(E_{l,{{{\mu}}}})\; \frac{e^{-it
E_{l,{{{\mu}}}}}}{\sks{E_{l,{{{\mu}}}}}{E_{l,{{{\mu}}}}}} ,
 \eeq
where $\mu=(r_0,r_1,N)$ and $\nu=(q_0,q_1,S),$ are expressed in terms of the orthogonal polynomials.
 From (\ref{0.5}) and  (\ref{2.7}) the time evolution of the expectation value
of any quantum observable $\mathbf X$ in a normalized state $\mket{\psi} \in {\cal H}$ becomes
\begin{multline}
  \langle \mathbf X(t) \rangle_{\psi} \equiv  \mbra{\psi}  e^{i\mathbf H_I \,t}  e^{i\mathbf H_0 \,t} \mathbf X
 e^{-i\mathbf H_0 \,t} e^{-i\mathbf H_I \,t} \mket{\psi} = \sum_{\mu,\nu\in J}^\infty
\sum_{m,r,l=0}^N\sum_{k,s,n=0}^S \sks{\psi}{m}_\mu \;_\mu\mbra{r}\mathbf X|s\rangle_{\nu}\;_{\nu}\langle n|
\psi \rangle \times\\\times e^{-it\Big(\omega_0(q_0-r_0)+\omega_1(q_1-r_1)+\omega_1k_1(S-N)+(\omega_0k_0-\omega_1k_1)(s-r)\Big)}\times\\
\times \frac{e^{-it\big(E_{k,\nu} - E_{l,\mu}\big) }}{\sks{E_{l,\mu}}{E_{l,\mu}} \sks{E_{k,\nu}}{E_{k,\nu}}}\;
\;\overline{P_{m}^\mu(E_{l,\mu})}\;\overline{ P_{s}^\nu(E_{k,\nu})}
P_{r}^\mu\!(E_{l,\mu})\;P_{n}^\nu\!(E_{k,\nu}).\lbl{Bz}
\end{multline}
 Similarly matrix elements of $\mathbf X(t)$ are equal
 \begin{multline}
  \; _\mu\mbra{m}  \mathbf X (t) \mket{n}_\nu =
\sum_{l,r=0}^N \sum_{k,s=0}^S \;_\mu\mbra{r}\mathbf X\mket{s}_\nu
 \;\,e^{-it\Big(\omega_0(q_0-r_0)+\omega_1(q_1-r_1)+\omega_1k_1(S-N)+(\omega_0k_0-\omega_1k_1)(s-r)\Big)}\times\\
 \times \frac{e^{-it
 \big(E_{k,\nu} - E_{l,\mu}\big)
}}{\sks{E_{l,\mu}}{E_{l,\mu}} \sks{E_{k,\nu}}{E_{k,\nu}}}\; \;\overline{P_{m}^\mu(E_{l,\mu})}\;\overline{
P_{s}^\nu(E_{k,\nu})} P_{r}^\mu\!(E_{l,\mu})\;P_{n}^\nu\!(E_{k,\nu}). \lbl{Bzz}
\end{multline}
\section{Hamiltonians related to the same families of fixed orthonormal polynomials }
\renewcommand{\theequation}{4.\arabic{equation}}
\setcounter{equation}{0}

It is natural that different Hamiltonians after reduction can lead to the same family of orthogonal polynomials. In this
section we solve the inverse problem, i.e. how to construct  different Hamiltonians which are related to the same
family of orthogonal polynomials.

Let us consider the interaction Hamiltonian in the case when $k_0=k_1=1$
  \beq
\mathbf H_I=h_0\left( \mathbf a_0^{*}\mathbf a_0,\mathbf a_1^{*} \mathbf a_1\right) + g\left( \mathbf a_0^{*}\mathbf
a_0,\mathbf a_1^{*}\mathbf a_1\right) \mathbf a_0  \mathbf a_1^{*}+ \mathbf a_0^{*}\mathbf a_1\overline{g}\left( \mathbf
a_0^{*}\mathbf a_0, \mathbf a_1^{*}\mathbf a_1\right)\lbl{Xa}.
      \eeq
Here the multi--index $\mu$  describing the decomposition (\ref{0.13}) is  a  single
       index $\mu\equiv N\in\mathbb{N}\cup\{0\}$.
 Moreover,   all components of that
 decomposition have different dimensions. Let us assume that  the eigenproblem is solved by a family of orthonormal finite
polynomials $\{P_n^N(E_{l,N})\}_{n=0}^N$. We will call (\ref{Xa}) the initial Hamiltonian.

 For any fixed $k_0,\;k_1\in\mathbb{N}$ we construct the Hamiltonian
  \beq
\widetilde{\mathbf H}_I=\widetilde{h}_0\left( \mathbf a_0^{*}\mathbf a_0,\mathbf a_1^{*} \mathbf a_1\right) +
\widetilde{g}\left( \mathbf a_0^{*}\mathbf a_0,\mathbf a_1^{*}\mathbf a_1\right) \mathbf a_0 ^{k_0} \left(\mathbf
a_1^{*} \right)^{k_1}+\left( \mathbf a_0^{*}\right)^{k_0}\mathbf a_1^{k_1}\overline{\,\widetilde{g}\,}\left( \mathbf
a_0^{*}\mathbf a_0, \mathbf a_1^{*}\mathbf a_1\right)\lbl{Xb}
      \eeq
    in  such a way that  the eigenproblem for any $\widetilde{\mathbf
       H}_\mu$, where $\mu=(r_0,r_1,N)\in J=
       \{0,1,\ldots,k_0-1\}\times\{0,1,\ldots,k_1-1\}\times(\mathbb{N}\cup\{0\})$,
       is solved by the initial family $\{P_n^N(E_{l,N})\}_{n=0}^N$, i.e.
 \beq
  \forall \mu=(r_0,r_1,N)\in J\quad\quad P_n^\mu=P_n^N \;\;\textrm{and}
  \;\;\;E_{l,\mu}=E_{l,N},\quad\quad n,l=0,1,\ldots,N.\lbl{3.3}
   \eeq
    In other words, the solution of eigenproblem for $\mathbf
    H_\mu$ does not depend on $r_0$ and $r_1$.

\;

       Let us introduce the following operator
      \begin{multline}
  W_{k_0k_1}(\mathbf a_0^*\mathbf a_0^{},\mathbf a_1^*\mathbf
  a_1^{}):=\sqrt{(\mathbf a_0^*\mathbf a_0^{}-\mathbf R_0+k_0)(\mathbf a_1^*\mathbf a_1^{}-\mathbf
  R_1)}\times\\
  \times\frac{1}{\sqrt{k_0k_1(\mathbf a_0^*\mathbf a_0^{}+1)(\mathbf a_0^*\mathbf a_0^{}+2)\ldots(\mathbf a_0^*\mathbf a_0^{}+k_0)
  \mathbf a_1^*\mathbf a_1^{}(\mathbf a_1^*\mathbf a_1^{}-1)\ldots(\mathbf a_1^*\mathbf
  a_1^{}-k_1+1)}}.
\end{multline}
 In particular $ W_{11}(\mathbf a_0^*\mathbf
a_0^{},\mathbf a_1^*\mathbf
  a_1^{})\equiv\textbf{1}$. It is easy to check that
  \beq
   W_{k_0k_1}(\mathbf a_0^*\mathbf a_0^{},\mathbf a_1^*\mathbf
  a_1^{})\mathbf a_0^{k_0}(\mathbf a_1^*)^{k_1}
  \mket{n}_\mu=\sqrt{n(N-n+1)}\mket{n-1}_\mu.
\eeq Since additionally we have that
 \beq
  &\frac{1}{k_0}(\mathbf a_0^*\mathbf
a_0^{}-\mathbf
R_0)\mket{n}_\mu=n\mket{n}_\mu&\\
&\frac{1}{k_1} (\mathbf a_1^*\mathbf a_1^{}-\mathbf
  R_1)\mket{n}_\mu=(N-n)\mket{n}_\mu,&
\eeq then putting
 \beq
 \widetilde{g}\left( \mathbf a_0^{*}\mathbf a_0,\mathbf
a_1^{*}\mathbf a_1\right) =g\left( \frac{1}{k_0}(\mathbf a_0^*\mathbf a_0^{}-\mathbf R_0),\frac{1}{k_1} (\mathbf
a_1^*\mathbf a_1^{}-\mathbf
  R_1)\right) W_{k_0k_1}(\mathbf a_0^*\mathbf a_0^{},\mathbf a_1^*\mathbf
  a_1^{})
   \eeq
       and
     \beq
\widetilde{h}_0\left( \mathbf a_0^{*}\mathbf a_0,\mathbf a_1^{*} \mathbf a_1\right)=h_0\left( \frac{1}{k_0}(\mathbf
a_0^*\mathbf a_0^{}-\mathbf R_0),\frac{1}{k_1} (\mathbf a_1^*\mathbf a_1^{}-\mathbf
  R_1)\right),
     \eeq
     we obtain the Hamiltonian (\ref{Xb}) which satisfies the conditions
      (\ref{3.3}).

     In the next section we present some examples of initial
     Hamiltonians which are related to the families of the finite discrete orthonormal
     polynomials, which are best known in the
     literature.

\section{Quantum systems related to some classes of finite orthonormal polynomials}
\renewcommand{\theequation}{5.\arabic{equation}}
\setcounter{equation}{0}

In this section  we present a list of Hamiltonians for which the spectral decompositions can be expressed by  some
selected families of finite orthogonal polynomials. The results of the previous section allows to restrict our list to
the cases when $k_{0} = k_{1} = 1.$  The notation used in this section is the same
 as in \cite{K-S}. The hypergeometric series are denoted by
  \beq
_rF_s\left(\begin{array}{c} a_1,\ldots,a_r\\
                         b_1,\ldots ,b_s
            \end{array}\vline \; z
      \right):=\sum_{n=0}^\infty\frac{(a_1,\ldots,a_r)_n}{(b_1,\ldots,b_s)_n}\frac{z^n}{n!},
    \eeq
where $(a_1,\ldots,a_r)_n:=(a_1)_n\cdots(a_r)_n$,\;
$(a)_n:=a(a+1)(a+2)\cdots(a+n-1) $ for~$n=1,2,\ldots$ and
$(a)_0:=1$. The basic hypergeometric series is defined by
 \beq
 _r\phi_s\left(\begin{array}{c}
 a_1,\ldots,a_r\\
                         b_1,\ldots ,b_s
            \end{array}\vline \;q;\; z
      \right):=\sum_{n=0}^\infty\frac{(a_1,\ldots,a_r;q)_n}{(b_1,\ldots,b_s;q)_n}\left(-1\right)^{(1+s-r)n}q^{(1+s-r)\left(
      ^n_2\right)}\frac{z^n}{(q;q)_n}
  \eeq
for $0<q<1,$ where
$(a_1,\ldots,a_r;q)_n:=(a_1;q)_n\cdots(a_r;q)_n$
and~$(a;q)_n:=(1-a)(1-aq)(1-aq^2)\cdots(1-aq^{n-1}),$ for~
$n=1,2,\ldots $ and $(a;q)_0=1$.

\subsection{Integrable systems related  to the Krawtchouk polynomials}

The Krawtchouk polynomials arise  for a system described by the Hamiltonian
\begin{eqnarray}
  \mathbf H_I=p\;\mathbf a_1^{*} \mathbf a_1+(1-p)\;\mathbf a_0^{*}\mathbf a_0 +\sqrt{p(1-p)} (\;\mathbf a_0 \mathbf a_1^{*}+
 \mathbf a_0^{*} \mathbf a_1^{}), \lbl{Ca}
\end{eqnarray}
where $0<p<1$. The spectrum of $\mathbf H_I$  is
 \beq
 \sigma(\mathbf H_I)=\mathbb{N}\;\cup\{0\},
   \eeq
   which follows from the fact that the eigenvalues of the reduced Hamiltonian $\mathbf H_N$ are $E_{l,N}=l$ for each $N$. Thus the
   eigenspaces ${\mathcal H}^l$, $l\in \sigma(\mathbf H_I)$ are the infinite--dimensional Hilbert subspaces
     \beq
{\mathcal H}^l =\textrm{span}\Big\{|E_{l,N}\rangle:\;N=l+n,\;n=0,1,\ldots\Big\},
   \eeq
   where
   \beq
    \mket{E_{l,N}}=\sum\limits_{n=0}^NK_n(E_{l,N};p,N)\mket{n,N-n}
     \eeq
and
\beq
K_n(E_{l,N};p,N)=\sqrt{ \frac{(-N)_n}{(-1)^nn!}\left(\frac{p}{1-p}\right)^n}\,_2F_1\left(\begin{array}{c}-n,-l \\
                                                 -N \
                                         \end{array}\vline\;\frac{1}{p}\right)
                                          \eeq
  are the Krawtchouk polynomials. For each   fixed $N$,  due to (\ref{Bd}), the
   finite family $\Big\{K_n(E_{l,N};p,N)\Big\}_{n=0}^N$
 forms an orthonormal system   with respect to the weight function
 \beq
  \frac{1}{\langle E_{l,N} | E_{l,N}
 \rangle}=\left(\begin{array}{c}
                                N \\
                                  l \
                                         \end{array}\right)p^l(1-p)^{N-l}.
                                           \eeq
We can summarize that the spectral decomposition of the interaction
Hamiltonian is
\begin{eqnarray}
\mathbf H_{I} = \sum_{N=0}^{\infty} \sum_{l=0}^{N} l\;\;\frac{\mket{E_{l,N}}\mbra{E_{l,N}}}{\sks{E_{l,N}}{E_{l,N}}}.
\end{eqnarray}

The Hamiltonian (\ref{Ca}) is quadratic in annihilation and creation operators and therefore the system is also
integrable via Heisenberg equations.
\subsection{Integrable systems related  to the Dual Hahn polynomials}

The Dual Hahn polynomials are related to the system given by the Hamiltonian
\begin{eqnarray}
  \mathbf H_I&=& \mathbf a_0^{*}\mathbf a_0(\mathbf a_1^{*} \mathbf a_1+\delta+1)+(\mathbf a_0^{*}\mathbf a_0+
  \gamma+1)\mathbf a_1^{*} \mathbf a_1 +\lbl{Cb}\\
  &&
  + \sqrt{(\mathbf a_1^{*} \mathbf a_1+\delta)(\mathbf a_0^{*}\mathbf a_0+\gamma+1)}\mathbf a_0 \mathbf a_1^{*}
+  \mathbf a_0^{*} \mathbf a_1^{}\sqrt{(\mathbf a_1^{*} \mathbf
a_1+\delta)(\mathbf a_0^{*}\mathbf a_0+\gamma+1)}\nonumber
\end{eqnarray}
where $\gamma,\;\delta>-1$. The spectrum of this Hamiltonian   is
 \beq
 \sigma(\mathbf H_I)=\Big\{l(l+\gamma+\delta+1):l=0,1,\ldots\Big\}
   \eeq
  and the eigenvalues of the reduced Hamiltonian $\mathbf H_N$ are $E_{l,N}=l(l+\gamma+\delta+1)$  and do not
   depend on $N$. For each eigenvalue $l(l+\gamma+\delta+1)$ the corresponding eigenspace
    ${\mathcal H}^l$ of ${\mathbf H_I}$  is the infinite--dimensional Hilbert space
     \beq
{\mathcal H}^l=\textrm{span}\Big\{|E_{l,N}\rangle:\;N=l+n,n=0,1,\ldots\Big\}
   \eeq
   where
   \beq
    \mket{E_{l,N}}=\sum\limits_{n=0}^NR_n(E_{l,N};\gamma,\delta,N)\mket{n,N-n}
     \eeq
   and
   \beq R_n(E_{l,N};\gamma,\delta,N)=\sqrt{
 \left(^{\gamma+n}_n\right)\left(^{\delta+N-n}_{N-n}\right)}
  \,_3F_2\left(\begin{array}{c}
                                                -n,-l,l+\gamma+\delta+1 \\
                                                 \gamma+1,-N \
                                         \end{array}\vline\;1\right)
                                          \eeq
                                                are the Dual Hahn polynomials. The finite family $\Big\{ R_n(E_{l,N}
                                                ;\gamma,\delta,N)\Big\}_{n=0}^N$, due to (\ref{Bd}),  forms an orthonormal system
                                         with respect to the weight function
 \beq
  \frac{1}{\langle E_{l,N} | E_{l,N}
 \rangle}=\displaystyle{\frac{(2l+\gamma+\delta+1)(\gamma+1)_l
                    (-1)^l(-N)_l\;N!}{(l+\gamma+\delta+1) _{N+1}(\delta+1)_l\;l!}}.
                      \eeq

  We can summarize that the spectral decomposition of the interaction
Hamiltonian is
\begin{eqnarray}
\mathbf H_{I} = \sum_{N=0}^{\infty} \sum_{l=0}^{N}
l(l+\gamma+\delta+1)\;\;\frac{\mket{E_{l,N}}\mbra{E_{l,N}}}{\sks{E_{l,N}}{E_{l,N}}}.
\end{eqnarray}

\subsection{Integrable systems related  to the discrete Chebyshev polynomials}

The Hamiltonian
\begin{eqnarray}
  \mathbf H_I&=&\frac{ (2\mathbf a_0^{*}\mathbf a_0+\mathbf a_1^{*} \mathbf a_1+ 1)
  \mathbf a_0^{*}\mathbf a_0}{2 (2\mathbf a_0^{*}\mathbf a_0
            + 1)}+\frac{(\mathbf a_0^{*}\mathbf a_0+ 1)\mathbf a_1^{*}
            \mathbf a_1}{2(2\mathbf a_0^{*}\mathbf a_0+ 1)}+\lbl{Cc}\\
            &&+\frac{1}{2}\sqrt{\frac{(2\mathbf a_0^{*}\mathbf a_0+\mathbf a_1^{*} \mathbf a_1+2)(\mathbf a_0^{*}
            \mathbf a_0+ 1) }
            {(2\mathbf a_0^{*}\mathbf a_0+ 1)(2\mathbf a_0^{*}\mathbf a_0+ 3)}} \mathbf a_0 \mathbf a_1^{*}+
            \mathbf a_0^{*}
            \mathbf a_1^{}\frac{1}{2}\sqrt{\frac{(2\mathbf a_0^{*}\mathbf a_0+\mathbf a_1^{*} \mathbf a_1+2)
            (\mathbf a_0^{*}
            \mathbf a_0+ 1) }
            {(2\mathbf a_0^{*}\mathbf a_0+ 1)(2\mathbf a_0^{*}\mathbf a_0+ 3)}}\nonumber
\end{eqnarray}
describes the system for which the solution of the eigenproblem is given by discrete Chebyshev polynomials. The
spectrum  of (\ref{Cc}) is
\beq
 \sigma(\mathbf H_I)=\mathbb{ N}\cup\{0\}
  \eeq
  and similarly as in subsection 5.1,
 the infinite--dimensional  eigenspaces ${\mathcal H}^l$ of ${\mathbf H_I}$ are spanned by the vectors
  \beq
    \mket{E_{l,N}}=\sum\limits_{n=0}^NT_n(E_{l,N};N)\mket{n,N-n},\;\;\;\;\;\; N=l+n,\;n=0,1,\ldots
     \eeq
with the eigenvalues $E_{l,N}=l$,\; $l=0,1,\ldots$ of the reduced Hamiltonian $\mathbf H_N.$
The coefficients
 \beq
 T_n(E_{l,N};N)= \sqrt{\frac{(2n+1)(-N)_nN!}{(-1)^n(n+1)_{N+1}n!}}  \,_3F_2\left(\begin{array}{c}
                                                -n,n+1,-l, \\
                                                1,-N \
                                         \end{array}\vline\;1\right)
                                           \eeq
                                           are the discrete Chebyshev polynomials. The  weight function for the family
                                           $\Big\{T_n(E_{l,N};N)\Big\}_{n=0}^N$ is, as for classical Chebyshev polynomials,
                                             constant
 \beq
  \frac{1}{\langle E_{l,N} | E_{l,N}
    \rangle}\equiv1.
     \eeq
     We can summarize that the spectral decomposition of the interaction
Hamiltonian is
\begin{eqnarray}
\mathbf H_{I} = \sum_{N=0}^{\infty} \sum_{l=0}^{N}
l\;\;\mket{E_{l,N}}\mbra{E_{l,N}}.
\end{eqnarray}

\subsection{Integrable systems related  to the Hahn polynomials}

The Hahn polynomials arise for   the system described by the Hamiltonian
\begin{eqnarray}
  \mathbf H_I&=&\frac{\mathbf a_0^{*}\mathbf a_0(2\mathbf a_0^{*}\mathbf a_0+\mathbf a_1^{*} \mathbf a_1+\alpha+\beta+1)
       (\mathbf a_0^{*}\mathbf a_0+\beta)}{(2\mathbf a_0^{*}\mathbf a_0+\alpha+\beta)(2\mathbf a_0^{*}\mathbf a_0
            +\alpha+\beta+1)}+\lbl{Cd}\\
            &+&\frac{(\mathbf a_0^{*}\mathbf a_0+\alpha+1)(\mathbf a_0^{*}\mathbf a_0+\alpha+\beta+1)
             \mathbf a_1^{*} \mathbf a_1}{(2\mathbf a_0^{*}\mathbf a_0+\alpha+\beta+1)(2\mathbf a_0^{*}\mathbf a_0+
             \alpha+\beta+2)}+\nonumber\\
            &+&\!\!\!\!\sqrt{\frac{(2\mathbf a_0^{*}\mathbf a_0+\mathbf a_1^{*} \mathbf a_1+\alpha+\beta+2)
            (\mathbf a_0^{*}\mathbf a_0+\beta+1)(\mathbf a_0^{*}\mathbf a_0+\alpha+1)(\mathbf a_0^{*}
            \mathbf a_0+\alpha+\beta+1)}
            {(2\mathbf a_0^{*}\mathbf a_0+\alpha+\beta+1)(2\mathbf a_0^{*}\mathbf a_0+
            \alpha+\beta+2)^2(2\mathbf a_0^{*}\mathbf a_0+\alpha+\beta+3)}} \mathbf a_0 \mathbf a_1^{*}+\nonumber\\
           &+&\!\! \mathbf a_0^{*} \mathbf a_1\sqrt{\frac{(2\mathbf a_0^{*}\mathbf a_0+\mathbf a_1^{*} \mathbf a_1+
           \alpha+\beta+2)(\mathbf a_0^{*}\mathbf a_0+\beta+1)(\mathbf a_0^{*}\mathbf a_0+\alpha+1)(\mathbf a_0^{*}
           \mathbf a_0+\alpha+\beta+1)}
            {(2\mathbf a_0^{*}\mathbf a_0+\alpha+\beta+1)(2\mathbf a_0^{*}\mathbf a_0+
            \alpha+\beta+2)^2(2\mathbf a_0^{*}\mathbf a_0+\alpha+\beta+3)}}\nonumber
\end{eqnarray}
where $\alpha,\; \beta>-1$. The spectrum   of $\mathbf H_I$ is
 \beq
 \sigma(\mathbf H_I)=\mathbb{ N}\;\cup\{0\}.
   \eeq
   and the infinite dimensional eigenspaces are spanned  by the vectors
   \beq
    \mket{E_{l,N}}=\sum\limits_{n=0}^NQ_n(E_{l,N};\alpha,\beta,N)\mket{n,N-n},
    \;\;\;\;N=l+n,\;n=0,1,\ldots\;,
     \eeq
 where $E_{l,N}=l$   and
 \beq
Q_n(E_{l,N};\alpha,\beta,N)=\sqrt{ \frac{(2n+\alpha+\beta+1)(\alpha+1)_n(-N)_n\;N!}
  {(-1)^n(n+\alpha+\beta+1)_{N+1}
  (\beta+1)_n\;n!}}  \,_3F_2\left(\begin{array}{c}
                                                -n,n+\alpha+\beta+1,-l, \\
                                                \alpha+1,-N \
                                         \end{array}\vline\;1\right)
                                          \eeq
                                           are the Hahn polynomials. The weight function for   the  family $\Big
                                           \{Q_n(E_{l,N};\alpha,\beta,N)\Big\}_{n=0}^N$  is
 \beq
  \frac{1}{\langle E_{l,N} | E_{l,N}
    \rangle}=\left(\begin{array}{c}
                                \alpha+l \\
                                  l \
                                         \end{array}\right)
                                  \left(\begin{array}{c}
                              \beta+N-l \\
                                N-l \
                                        \end{array}\right).
      \eeq
      We can summarize that the spectral decomposition of the interaction
Hamiltonian is
\begin{eqnarray}
\mathbf H_{I} = \sum_{N=0}^{\infty} \sum_{l=0}^{N} l\;\;\frac{\mket{E_{l,N}}\mbra{E_{l,N}}}{\sks{E_{l,N}}{E_{l,N}}}.
\end{eqnarray}
     Let us note that putting $\alpha=\beta=0$ we obtain the discrete Chebyshev polynomials.



\subsection{Integrable systems related  to the  Dual $q$--Hahn polynomials}

For any fixed $0<q<1$ and $0<\gamma,\;\delta<q^{-1}$ such that
$0<\gamma\delta<q^{-1}$  the Hamiltonian
\begin{eqnarray}
  \mathbf H_I&=& 1+\gamma\delta q-\gamma q(1-q^{\mathbf a_0^{*}\mathbf a_0})(\delta-q^{-\mathbf a_1^{*}\mathbf a_1-1})-(1-q^{-\mathbf a_1^{*}\mathbf a_1})
  (1-\gamma q^{\mathbf a_0^{*}\mathbf a_0+1})\\
                      &+&\!\! \sqrt{\frac{ \gamma q(1-q^{-\mathbf a_1^{*}\mathbf a_1})(1-\gamma q^{\mathbf a_0^{*}\mathbf a_0+1})(1-q^{\mathbf a_0^{*}\mathbf a_0+1})
                (\delta-q^{-\mathbf a_1^{*}\mathbf a_1})}{(\mathbf a_0^{*}\mathbf a_0+1)\mathbf a_1^{*}\mathbf a_1}}
\mathbf a_0 \mathbf a_1^{*}+\nonumber\\
&+&  \mathbf a_0^{*} \mathbf a_1^{}  \sqrt{\frac{ \gamma q(1-q^{-\mathbf a_1^{*}\mathbf a_1})(1-\gamma q^{\mathbf a_0^{*}\mathbf a_0+1})(1-q^{\mathbf a_0^{*}\mathbf a_0+1})
                (\delta-q^{-\mathbf a_1^{*}\mathbf a_1})}{(\mathbf a_0^{*}\mathbf a_0+1)\mathbf a_1^{*}\mathbf a_1}}
\nonumber
\end{eqnarray}
has the   spectrum
 \beq
\sigma(\mathbf H_I)=\big\{q^{-l}+\gamma\delta
q^{l+1}:l=0,1,\ldots\big\}
   \eeq
and the eigenspace ${\mathcal H}^l$ corresponding to eigenvalue
\;\;$q^{-l}+\gamma\delta q^{l+1}\in\sigma(\mathbf H_I)$ is
     \beq
    {\mathcal H}^l=\textrm{span}\Big\{|E_{l,N}\rangle:\;N=l+n,\;n=0,1,\ldots\Big\}
   \eeq
 with
 \beq
  E_{l,N}=q^{-l}+\gamma\delta
   q^{l+1},
   \eeq
   and
   \beq
    \mket{E_{l,N}}=\sum\limits_{n=0}^N R_n(E_{l,N};\gamma,\delta,N|q )\mket{n,N-n}
     \eeq
and
  \beq
 R_n\Big(E_{l,N};\gamma,\delta,N|q \Big)= \sqrt{\frac{(\delta q;q)_N(\gamma q,q^{-N};q)_n(\gamma q)^N}{
  (\gamma\delta q^2;q)_N(q,\delta^{-1}q^{-N};q)_n
  (\gamma\delta q)^n}}\;_3\phi_2\left(\begin{array}{c}
  {q^{-n},q^{-l},\gamma\delta  q^{l+1}}\\
  {\gamma q,q^{-N}}\end{array}\vline\;q;q\right)
                                         \eeq
                                          which are the Dual $q$--Hahn  polynomials. For each
                                         fixed $N$ the  family
                                         $ \Big\{R_n\Big(E_{l,N};\gamma,\delta,N|q \Big)\Big\}_{n=0}^N$
                                          forms an orthonormal system
                                         with respect to the weight function
 \beq
  \frac{1}{\langle E_{l,N} | E_{l,N}    \rangle}= \frac{\left(\gamma q,\gamma\delta q,q^{-N};q\right)
 _l\left(1-\gamma\delta q^{2l+1}\right)}
 {\left(q,\gamma\delta q^{N+2},\delta q;q\right)_l \left( 1-\gamma\delta q\right)\left(-\gamma q\right)^l}
 q^{Nl- \left(^l_2\right)}.
      \eeq

      We can summarize that the spectral decomposition of the interaction
Hamiltonian is
\begin{eqnarray}
\mathbf H_{I} = \sum_{N=0}^{\infty} \sum_{l=0}^{N} (q^{-l}+\gamma\delta
q^{l+1})\;\;\frac{\mket{E_{l,N}}\mbra{E_{l,N}}}{\sks{E_{l,N}}{E_{l,N}}}.
\end{eqnarray}
\subsection{Integrable systems related  to the Affine $q$--Krawtchouk polynomials}

For any fixed $0<q<1$ and  $0<p<q^{-1}$ the Hamiltonian
\begin{eqnarray}
  \mathbf H_I&=&  1-\left[(1-q^{-\mathbf a_1^{*}\mathbf a_1})(1-pq^{\mathbf a_0^{*}\mathbf a_0+1})-pq^{-\mathbf a_1^{*}
                        \mathbf a_1}  (1-q^{\mathbf a_0^{*}\mathbf a_0})\right]\\
                  &+&\!\! \sqrt{\frac{-pq^{-\mathbf a_1^{*}\mathbf a_1+1}(1-q^{\mathbf a_0^{*}\mathbf a_0+1})
                  (1-q^{-\mathbf a_1^{*}\mathbf a_1})(1-pq^{\mathbf a_0^{*}\mathbf a_0+1}) }{(\mathbf a_0^{*}
                  \mathbf a_0+1)\mathbf a_1^{*}\mathbf a_1}}
                        \mathbf a_0 \mathbf a_1^{*}+\nonumber\\
&+&  \mathbf a_0^{*} \mathbf a_1^{}   \sqrt{\frac{-pq^{-\mathbf a_1^{*}\mathbf a_1+1}(1-q^{\mathbf a_0^{*}\mathbf a_0+1})
                  (1-q^{-\mathbf a_1^{*}\mathbf a_1})(1-pq^{\mathbf a_0^{*}\mathbf a_0+1}) }{(\mathbf a_0^{*}
                  \mathbf a_0+1)\mathbf a_1^{*}\mathbf a_1}}
\nonumber
\end{eqnarray}
 has the spectrum
 \beq
  \sigma(\mathbf H_I)=\{q^{-l}:l=0,1,\ldots\}.
   \eeq
     Eigenspaces ${\mathcal H}^l$,\;\;$q^{-l}\in\sigma(\mathbf H_I)$ are
     \beq
    {\mathcal H}^l=span\Big\{|E_{l,N}\rangle:\;N=l+n,\;n=0,1,\ldots\Big\}
   \eeq
  for  $ E_{l,N}=q^{-l} $,
   \beq
    \mket{E_{l,N}}=\sum\limits_{n=0}^N  K^{Aff}_n(E_{l,N};p,N;q )\mket{n,N-n}
     \eeq
  and
  \beq
   K^{Aff}_n(E_{l,N};p,N;q )= \sqrt{\frac{(pq)^{N-n}(pq;q)_n(q;q)_N}{(q;q)_n(q;q)_{N-n}}}\;_3\phi_2\left(\begin{array}{c}
    {q^{-n},0,q^{-l}}\\
    {pq,q^{-N}}\end{array}\vline\;q;q\right)
    \eeq
                                          which are the affine $q$--Krawtchouk polynomials. The   family
                                          $  \Big\{K^{Aff}_n(E_{l,N};p,N;q )\Big\}_{n=0}^N$
                                           is the orthonormal system
                                         with respect to the weight function
                                          \beq
  \frac{1}{\langle E_{l,N} | E_{l,N}    \rangle}= \frac{(pq;q)_{l}(q;q)_N}
  {(q;q)_l(q;q)_{N-l}}\left(pq\right)^{-l}.
      \eeq

      We can summarize that the spectral decomposition of the interaction
Hamiltonian is
\begin{eqnarray}
\mathbf H_{I} = \sum_{N=0}^{\infty} \sum_{l=0}^{N}
q^{-l}\;\;\frac{\mket{E_{l,N}}\mbra{E_{l,N}}}{\sks{E_{l,N}}{E_{l,N}}}.
\end{eqnarray}

\subsection{Integrable systems related  to the $q$--Krawtchouk polynomials}
The Hamiltonian
\begin{eqnarray}
  \mathbf H_I&=& 1-\frac{(1-q^{-\mathbf a_1^{*}\mathbf a_1})(1+pq^{\mathbf a_0^{*}\mathbf a_0})}{(1+pq^{2\mathbf a_0^{*}\mathbf a_0})(1+pq^{2\mathbf a_0^{*}\mathbf a_0+1})}+pq^{\mathbf a_0^{*}\mathbf a_0-\mathbf a_1^{*}\mathbf a_1-1}
                  \frac{(1+pq^{2\mathbf a_0^{*}\mathbf a_0+\mathbf a_1^{*}\mathbf a_1})
                  (1-q^{\mathbf a_0^{*}\mathbf a_0})}{(1+pq^{2\mathbf a_0^{*}\mathbf a_0-1})
                  (1+pq^{2\mathbf a_0^{*}\mathbf a_0})} + \\
                        &+&\!\! \sqrt{-\frac{pq^{\mathbf a_0^{*}\mathbf a_0-\mathbf a_1^{*}\mathbf a_1+1}(1+pq^{2\mathbf a_0^{*}\mathbf a_0+\mathbf a_1^{*}\mathbf a_1+1})(1-q^{\mathbf a_0^{*}\mathbf a_0+1})
                    (1-q^{-\mathbf a_1^{*}\mathbf a_1})(1+pq^{\mathbf a_0^{*}\mathbf a_0})}{(1+pq^{2\mathbf a_0^{*}\mathbf a_0})
                  (1+pq^{2\mathbf a_0^{*}\mathbf a_0+1})^2   (1+pq^{2\mathbf a_0^{*}\mathbf a_0+2}){(\mathbf a_0^{*}
                  \mathbf a_0+1)\mathbf a_1^{*}\mathbf a_1}               }}
\mathbf a_0 \mathbf a_1^{*}+\nonumber\\
&+&  \mathbf a_0^{*} \mathbf a_1^{}    \sqrt{-\frac{pq^{\mathbf a_0^{*}\mathbf a_0-\mathbf a_1^{*}\mathbf a_1+1}(1+pq^{2\mathbf a_0^{*}\mathbf a_0+\mathbf a_1^{*}\mathbf a_1+1})(1-q^{\mathbf a_0^{*}\mathbf a_0+1})
                    (1-q^{-\mathbf a_1^{*}\mathbf a_1})(1+pq^{\mathbf a_0^{*}\mathbf a_0})}{(1+pq^{2\mathbf a_0^{*}\mathbf a_0})
                  (1+pq^{2\mathbf a_0^{*}\mathbf a_0+1})^2   (1+pq^{2\mathbf a_0^{*}\mathbf a_0+2}){(\mathbf a_0^{*}
                  \mathbf a_0+1)\mathbf a_1^{*}\mathbf a_1}               }}\nonumber,
\end{eqnarray}
  where $0<q<1$ and $p>0$, has the   spectrum
 \beq
  \sigma(\mathbf H_I)=\{q^{-l}:l=0,1,\ldots\}.
   \eeq
     Eigenspaces ${\mathcal H}^l$,\;\;$q^{-l}\in\sigma(\mathbf H_I)$ are
     \beq
    {\mathcal
    H}^l=\textrm{span}\Big\{|E_{l,N}\rangle:\;N=l+n,n=0,1,\ldots\Big\},
   \eeq
   where for $ E_{l,N}=q^{-l} $
   \beq
    \mket{E_{l,N}}=\sum\limits_{n=0}^N   K_n(E_{l,N};p,N;q )\mket{n,N-n}
     \eeq
     and
  \beq
   K_n(E_{l,N};p,N;q )=\sqrt{
  \frac{(-p,q^{-N};q)_n(1+pq^{2n})p^Nq^{\left(^{N+1}_{\;\;\;2}\right)}}
  {(q,-pq^{N+1};q)_n(1+p)(-pq;q)_N(-pq^{-N})^nq^{n^2}} }\;_3\phi_2\left(\!\!\!\!\begin{array}{c}
  {q^{-n},q^{-l},-pq^n}\\
  {q^{-N},0}\end{array}\vline \;q;q\right)
  \eeq
                                           are the $q$--Krawtchouk polynomials.
                                           The weight function for the orthonormal system
                                            $ \Big\{K_n(E_{l,N};p,N;q )\Big\}^N_{n=0}$
                                          is
  \beq
  \frac{1}{\langle E_{l,N} | E_{l,N}    \rangle}= \frac{(q^{-N};q)_{l}}{(q;q)_l}\left(-p\right)^{-l}.
      \eeq

We can summarize that the spectral decomposition of the interaction
Hamiltonian is
\begin{eqnarray}
\mathbf H_{I} = \sum_{N=0}^{\infty} \sum_{l=0}^{N}
q^{-l}\;\;\frac{\mket{E_{l,N}}\mbra{E_{l,N}}}{\sks{E_{l,N}}{E_{l,N}}}.
\end{eqnarray}
\subsection{Integrable systems related  to the  $q$--Hahn polynomials}

The Hamiltonian
\begin{eqnarray}
  \mathbf H_I&=& \nonumber 1+\frac{\alpha q^{-\mathbf a_1^{*}\mathbf a_1}(1-q^{\mathbf a_0^{*}\mathbf a_0})(1-\alpha\beta q^{2\mathbf a_0^{*}\mathbf a_0+\mathbf a_1^{*}\mathbf a_1+1})
             (1-\beta q^{\mathbf a_0^{*}\mathbf a_0})}
            {(1-\alpha\beta q^{2\mathbf a_0^{*}\mathbf a_0})(1-\alpha\beta q^{2\mathbf a_0^{*}\mathbf a_0+1})}- \\
            \nonumber &-&\frac{(1- q^{-\mathbf a_1^{*}\mathbf a_1})(1-\alpha q^{\mathbf a_0^{*}\mathbf a_0+1})
            (1-\alpha\beta q^{\mathbf a_0^{*}\mathbf a_0+1})}
             {(1-\alpha\beta q^{2\mathbf a_0^{*}\mathbf a_0+2})(1-\alpha\beta q^{2\mathbf a_0^{*}\mathbf a_0+1})}+\\
                &+&\! \sqrt{-\frac{\alpha q^{-\mathbf a_1^{*}\mathbf a_1+1}(1-q^{\mathbf a_0^{*}\mathbf a_0+1})
                (1-\alpha\beta q^{2\mathbf a_0^{*}\mathbf a_0+\mathbf a_1^{*}\mathbf a_1+2})
            (1-\beta q^{\mathbf a_0^{*}\mathbf a_0+1})}{(1-\alpha\beta
                q^{2\mathbf a_0^{*}\mathbf a_0+1})(1-\alpha\beta q^{2\mathbf a_0^{*}\mathbf a_0+2})^2(1-\alpha\beta
              q^{2\mathbf a_0^{*}\mathbf a_0+3})(\mathbf a_0^{*}\mathbf a_0+1)\mathbf a_1^{*}\mathbf a_1}}\times\\
      &\times&\sqrt{ (1- q^{-\mathbf a_1^{*}\mathbf a_1})(1-\alpha q^{\mathbf a_0^{*}\mathbf a_0+1})(1-\alpha\beta
      q^{\mathbf a_0^{*}\mathbf a_0+1})}\mathbf a_0 \mathbf a_1^{*}+\nonumber\\
&+&  \mathbf a_0^{*} \mathbf a_1^{}   \sqrt{-\frac{\alpha q^{-\mathbf a_1^{*}\mathbf a_1+1}(1-q^{\mathbf a_0^{*}
\mathbf a_0+1})(1-\alpha\beta q^{2\mathbf a_0^{*}\mathbf a_0+\mathbf a_1^{*}\mathbf a_1+2})
            (1-\beta q^{\mathbf a_0^{*}\mathbf a_0+1})}{(1-\alpha\beta
                q^{2\mathbf a_0^{*}\mathbf a_0+1})(1-\alpha\beta q^{2\mathbf a_0^{*}\mathbf a_0+2})^2(1-\alpha\beta
              q^{2\mathbf a_0^{*}\mathbf a_0+3})(\mathbf a_0^{*}\mathbf a_0+1)\mathbf a_1^{*}\mathbf a_1}}\times\nonumber\\
     &\times&         \sqrt{ (1- q^{-\mathbf a_1^{*}\mathbf a_1})(1-\alpha q^{\mathbf a_0^{*}\mathbf a_0+1})(1-\alpha\beta
              q^{\mathbf a_0^{*}\mathbf a_0+1})},\nonumber
\nonumber
\end{eqnarray}
 where  $0<q<1$,  $0<\alpha<q^{-1}$ and $0<\beta<q^{-1}$ has  the   spectrum
 \beq
  \sigma(\mathbf H_I)=\{q^{-l}:l=0,1,\ldots\}.
   \eeq
     Eigenspaces ${\mathcal H}^l$,\;\;$q^{-l}\in\sigma(\mathbf H_I)$ are
     \beq
    {\mathcal H}^l=\textrm{span}\Big\{|E_{l,N}\rangle:\;N=l+n,n=0,1,\ldots\Big\}
   \eeq
   where  $ E_{l,N}=q^{-l} $
   \beq
    \mket{E_{l,N}}=\sum\limits_{n=0}^N   Q_n(E_{l,N};\alpha,\beta,N|q)\mket{n,N-n}
     \eeq
 and
   \beq
   Q_n\big(E_{l,N};\alpha,\beta,N|q\big)&=&\\\nonumber
   &&\!\!\!\!\!\!\!\!\!\!\!\!\!\!\!\!\!\!\!\!\!\!\!\!\!\!\!\!\!\!\!\!\!\!\!\!\!\!\!\!\!\!\!\!\!\!\!\!\!\!\!\!
   =\sqrt{\frac{(\beta q;q)_N(\alpha q)^N}{(\alpha\beta q^2;q)_N}
  \frac{(\alpha q,\alpha\beta q,q^{-N};q)_n(1-\alpha\beta q^{2n+1})q^{Nn-\left(^{n}_{2}\right)}}{(q,\alpha\beta q^{N+2},
  \beta q ;q)_n(1-\alpha\beta q )(-\alpha q)^n
  }}
 \;_3\phi_2\left(\begin{array}{c}
  {q^{-n},\alpha\beta q^{n+1},q^{-l}}\\
  {\alpha q,q^{-N}}
  \end{array}\vline\;q;q\right)
   \eeq
                                           are the $q$--Hahn polynomials. The appropriate weight function is
 \beq
  \frac{1}{\langle E_{l,N} | E_{l,N}    \rangle}=\frac{\left(\alpha q,q^{-N};q\right)_l}{\left(q,\beta^{-1}q^{-N};q\right)_l}
 \left(\alpha\beta q\right)^{-l}.
      \eeq

We can summarize that the spectral decomposition of the interaction
Hamiltonian is
\begin{eqnarray}
\mathbf H_{I} = \sum_{N=0}^{\infty} \sum_{l=0}^{N}
q^{-l}\;\;\frac{\mket{E_{l,N}}\mbra{E_{l,N}}}{\sks{E_{l,N}}{E_{l,N}}}.
\end{eqnarray}
\subsection{Integrable systems related  to the Dual $q$--Krawtchouk polynomials}

The Hamiltonian
\begin{eqnarray}
  \mathbf H_I&=& (1+c)q^{-\mathbf a_1^{*}\mathbf a_1}+\nonumber\\
&+&
 \sqrt{\frac{cq^{-(\mathbf a_0^{*}\mathbf a_0+\mathbf a_1^{*}\mathbf a_1)}(1-q^{-\mathbf a_1^{*}\mathbf a_1})
 (1-q^{\mathbf a_0^{*}\mathbf a_0+1})}{(\mathbf a_0^{*}\mathbf a_0+1)\mathbf a_1^{*}\mathbf a_1}}
  \mathbf a_0^{} \mathbf a_1^{*}   +\\
&+&  \mathbf a_0^{*} \mathbf a_1^{} \sqrt{\frac{cq^{-(\mathbf a_0^{*}\mathbf a_0+\mathbf a_1^{*}\mathbf a_1)}
(1-q^{-\mathbf a_1^{*}\mathbf a_1})\nonumber
 (1-q^{\mathbf a_0^{*}\mathbf a_0+1})}{(\mathbf a_0^{*}\mathbf a_0+1)\mathbf a_1^{*}\mathbf a_1}},
\end{eqnarray}
where $0<q<1$, $c<0$, has the   spectrum
 \beq
  \sigma(\mathbf H_I)=\{q^{-l}+cq^{l-N}:l=0,1,\ldots N,\;N=0,1,\ldots\}.
   \eeq
The eigenvalues $E_{l,N}=q^{-l}+cq^{l-N} $, $l=0,1,\ldots,N$
of the reduced Hamiltonian $\mathbf H_N$  depend on $N$, and
therefore the eigenspaces of $\mathbf H_I$ are one dimensional
given by the vectors
   \beq
    \mket{E_{l,N}}=\sum\limits_{n=0}^N
    K_n\big(E_{l,N};c,N|q\big)\;\;\mket{n,N-n},
     \eeq
where the coefficients
  \beq
   K_n\big(E_{l,N};c,N|q\big)=\sqrt{
    \frac{(q^{-N};q)_n}{(c^{-1};q)_N(q;q)_n(cq^{-N})^n}}\;
      _3\phi_2\left(\begin{array}{c}
      {q^{-n},q^{-l},cq^{l-N}}\\
      {q^{-N},0}  \end{array}\vline\;q;q\right)
  \eeq
                                           are the dual  $q$--Krawtchouk  polynomials. For each
                                         fixed $N$ the finite family $ \Big\{K_n(E_{l,N};c,N|q)\Big\}^N_{n=0}$
                                          is an orthonormal system
                                        with respect to the weight function
 \beq
  \frac{1}{\langle E_{l,N} | E_{l,N}\rangle}= \frac{(cq^{-N},q^{-N};q)_{l}
    \left(1-cq^{2l-N}\right)}{(q,cq;q)_l
    \left(1-cq^{-N}\right) }c^{-l}q^{l(2N-l)}.
      \eeq
We can summarize that the spectral decomposition of the interaction
Hamiltonian is
\begin{eqnarray}
\mathbf H_{I} = \sum_{N=0}^{\infty} \sum_{l=0}^{N}
(q^{-l}+cq^{l-N})\;\;\frac{\mket{E_{l,N}}\mbra{E_{l,N}}}{\sks{E_{l,N}}{E_{l,N}}}.
\end{eqnarray}

 \section*{Acknowledgements}
 We would like to thank I. Jex
for fruitful discussion and J.Tolar for his interest in our work. The authors also acknowledge partial support of the
Ministry of Education of Czech Republic under the research project MSM210000018 and of the Czech Grant Agency grant No.
200/01/0318.

\end{document}